\documentclass[11pt]{article}
\usepackage{graphicx}
\usepackage{amsmath}
\usepackage{amssymb}
\usepackage{caption2}
\begin{document}
\bibliographystyle{prsty}
\begin{center}
{\large {\bf \sc{  Analysis of  the $Y(4140)$ with  QCD sum rules }}} \\[2mm]
Zhi-Gang Wang \footnote{E-mail,wangzgyiti@yahoo.com.cn.  }     \\
 Department of Physics, North China Electric Power University,
Baoding 071003, P. R. China
\end{center}

\begin{abstract}
In this article, we assume that there exists a scalar $D_s^\ast
{\bar D}_s^\ast$ molecular state in the $J/\psi \phi$ invariant mass
distribution, and study its mass using the QCD sum rules. The
predictions depend heavily on  the two criteria (pole dominance and
convergence of the operator product expansion) of the QCD sum rules.
The  value of the mass is about   $M_{D_s^\ast {\bar
D}_s^\ast}=(4.43\pm0.16)\,\rm{GeV}$, which
 is  inconsistent with the experimental data.
    The $D_s^\ast {\bar
D}_s^\ast$  is probably  a virtual state and not related to the
meson $Y(4140)$. Other possibility, such as a hybrid charmonium is
not excluded.
\end{abstract}



\section{Introduction}

Recently the CDF Collaboration   observed   a narrow structure
($Y(4140)$) near the $J/\psi\phi$ threshold with statistical
significance in excess of 3.8 standard deviations in exclusive
$B^+\to J/\psi\phi K^+$ decays produced in $\bar{p} p $ collisions
at $\sqrt{s}=1.96 \,\rm{TeV}$ \cite{CDF0903}. The mass and  width of
the structure are measured to be $4143.0\pm2.9\pm1.2\,\rm{ MeV}$ and
$11.7^{+8.3}_{-5.0}\pm3.7\, \rm{MeV}$, respectively. The meson
$Y(4140)$ is very similar to the charmonium-like state $Y(3930)$
near the $J/\psi \omega$ threshold \cite{Belle2005,Babar2008}. The
mass and width of the $Y(3930)$ are $3914.6^{+3.8}_{-3.4}\pm 2.0
\,\rm{MeV}$ and $34^{+12}_{-8}\pm 5\,\rm{MeV}$,  respectively
\cite{Babar2008}.

In Ref.\cite{Zhu0903}, Liu et al study the narrow structure
$Y(4140)$ with the meson-exchange model, and draw the conclusion
that the $Y(4140)$ is probably a $D_s^\ast {\bar D}_s^\ast$
molecular state with $J^{PC}=0^{++}$ or $2^{++}$ while the $Y(3930)$
is its $D^\ast {\bar D}^\ast$ molecular partner. In
Ref.\cite{Mahajan0903}, Mahajan argues that it is  likely to be a
$D_s^*\bar{D}_s^*$ molecular state or an exotic ($J^{PC}=1^{-+}$)
hybrid charmonium.

The mass is a fundamental parameter in describing a hadron, in order
to identify  the $Y(4140)$ as a scalar molecular state, we must
prove that its mass  lies  in the region $(4.1-4.2)\, \rm{GeV}$. In
this article, we assume that there exists  a scalar
$D_s^*\bar{D}_s^*$ molecular state in the $J/\psi\phi$ invariant
mass distribution, and study its mass with the QCD sum rules
\cite{SVZ79,Reinders85}.

 In the QCD sum rules, the operator product expansion is used to expand
the time-ordered currents into a series of quark and gluon
condensates which parameterize the long distance properties of  the
QCD vacuum. Based on the quark-hadron duality, we can obtain copious
information about the hadronic parameters at the phenomenological
side \cite{SVZ79,Reinders85}.

The article is arranged as follows:  we derive the QCD sum rules for
the mass of  the $Y(4140)$  in section 2; in section 3, numerical
results and discussions; section 4 is reserved for conclusion.

\section{QCD sum rules for  the molecular state $Y(4140)$ }
In the following, we write down  the two-point correlation function
$\Pi(p)$  in the QCD sum rules,
\begin{eqnarray}
\Pi(p)&=&i\int d^4x e^{ip \cdot x} \langle
0|T\left\{J(x)J^{\dagger}(0)\right\}|0\rangle \, , \\
J(x)&=&\bar{c}(x)\gamma_\mu s(x) \bar{s}(x)\gamma^\mu c(x) \, ,
\end{eqnarray}
we choose  the scalar current $J(x)$ to interpolate the molecular
state $Y(4140)$.

We can insert  a complete set of intermediate hadronic states with
the same quantum numbers as the current operator $J(x)$ into the
correlation function $\Pi(p)$  to obtain the hadronic representation
\cite{SVZ79,Reinders85}. After isolating the ground state
contribution from the pole term of the $Y(4140)$, we get the
following result,
\begin{eqnarray}
\Pi(p)&=&\frac{\lambda_{Y}^2}{M_{Y}^2-p^2} +\cdots \, \, ,
\end{eqnarray}
where the pole residue (or coupling) $\lambda_Y$ is defined by
\begin{eqnarray}
\lambda_{Y} &=& \langle 0|J(0)|Y(p)\rangle \, .
\end{eqnarray}

 In the following, we briefly outline  the operator product
expansion for the correlation function $\Pi(p)$  in perturbative
QCD. The calculations are performed at  the large space-like
momentum region $p^2\ll 0$. We write down the "full" propagators
$S_{ij}(x)$ and $C_{ij}(x)$ of a massive quark in the presence of
the vacuum condensates firstly \cite{Reinders85},
\begin{eqnarray}
S_{ij}(x)&=& \frac{i\delta_{ij}\!\not\!{x}}{ 2\pi^2x^4}
-\frac{\delta_{ij}m_s}{4\pi^2x^2}-\frac{\delta_{ij}}{12}\langle
\bar{s}s\rangle +\frac{i\delta_{ij}}{48}m_s
\langle\bar{s}s\rangle\!\not\!{x}-\frac{\delta_{ij}x^2}{192}\langle \bar{s}g_s\sigma Gs\rangle\nonumber\\
 &&+\frac{i\delta_{ij}x^2}{1152}m_s\langle \bar{s}g_s\sigma
 Gs\rangle \!\not\!{x}-\frac{i}{32\pi^2x^2} G^{ij}_{\mu\nu} (\!\not\!{x}
\sigma^{\mu\nu}+\sigma^{\mu\nu} \!\not\!{x})  +\cdots \, ,
\end{eqnarray}
\begin{eqnarray}
C_{ij}(x)&=&\frac{i}{(2\pi)^4}\int d^4k e^{-ik \cdot x} \left\{
\frac{\delta_{ij}}{\!\not\!{k}-m_c}
-\frac{g_sG^{\alpha\beta}_{ij}}{4}\frac{\sigma_{\alpha\beta}(\!\not\!{k}+m_c)+(\!\not\!{k}+m_c)\sigma_{\alpha\beta}}{(k^2-m_c^2)^2}\right.\nonumber\\
&&\left.+\frac{\pi^2}{3} \langle \frac{\alpha_sGG}{\pi}\rangle
\delta_{ij}m_c \frac{k^2+m_c\!\not\!{k}}{(k^2-m_c^2)^4}
+\cdots\right\} \, ,
\end{eqnarray}
where $\langle \bar{s}g_s\sigma Gs\rangle=\langle
\bar{s}g_s\sigma_{\alpha\beta} G^{\alpha\beta}s\rangle$  and
$\langle \frac{\alpha_sGG}{\pi}\rangle=\langle
\frac{\alpha_sG_{\alpha\beta}G^{\alpha\beta}}{\pi}\rangle$, then
contract the quark fields in the correlation function $\Pi(p)$ with
Wick theorem, and obtain the result:
\begin{eqnarray}
\Pi(p)&=&i\int d^4x e^{ip \cdot x}   Tr\left[\gamma_\mu
S_{ij}(x)\gamma_\alpha C_{ji}(-x) \right] Tr\left[\gamma^\mu
C_{mn}(x)\gamma^\alpha S_{nm}(-x) \right] \, ,
\end{eqnarray}
where the $i$, $j$, $m$ and $n$ are color indexes.

Substitute the full  $s$ and $c$ quark propagators into the
correlation function  $\Pi(p)$ and complete  the integral in the
coordinate space, then integrate over the variables in the momentum
space, we can obtain the correlation function $\Pi(p)$ at the level
of the quark-gluon degrees  of freedom.

We carry out the operator product expansion to the vacuum
condensates adding up to dimension-10 and
 take the assumption of vacuum saturation for the  high
dimension vacuum condensates, they  are always
 factorized to lower condensates with vacuum saturation in the QCD sum
 rules, and   factorization works well in  large $N_c$ limit.
 In calculation, we observe that the
contributions  from the gluon condensate are suppressed by large
denominators and would not play any significant roles
\cite{Wang1,Wang2,Wang0807,Wang0708}.

Once analytical results are obtained,   then we can take the
quark-hadron duality and perform Borel transform  with respect to
the variable $P^2=-p^2$, finally we obtain  the following sum rule:
\begin{eqnarray}
\lambda_{Y}^2 e^{-\frac{M_Y^2}{M^2}}= \int_{4(m_c+m_s)^2}^{s_0} ds
\rho(s) e^{-\frac{s}{M^2}} \, ,
\end{eqnarray}
where
\begin{eqnarray}
\rho(s)&=&\rho_{0}(s)+\rho_{\langle
\bar{s}s\rangle}(s)+\left[\rho^A_{\langle
GG\rangle}(s)+\rho^B_{\langle GG\rangle}(s)\right]\langle
\frac{\alpha_s GG}{\pi}\rangle+\rho_{\langle \bar{s}s\rangle^2}(s)
\, ,
\end{eqnarray}
the lengthy  expressions of the spectral densities $\rho_0(s)$,
$\rho_{\langle \bar{s}s\rangle}(s)$, $\rho^A_{\langle
GG\rangle}(s)$, $\rho^B_{\langle GG\rangle}(s)$ and $\rho_{\langle
\bar{s}s\rangle^2}(s)$ are presented in the appendix.

 Differentiating  the Eq.(8) with respect to  $\frac{1}{M^2}$, then eliminate the
 pole residue $\lambda_{Y}$, we can obtain a sum rule for
 the mass of the $Y(4140)$,
 \begin{eqnarray}
 M_Y^2= \frac{\int_{4(m_c+m_s)^2}^{s_0} ds
\frac{d}{d \left(-1/M^2\right)}\rho(s)e^{-\frac{s}{M^2}}
}{\int_{4(m_c+m_s)^2}^{s_0} ds \rho(s)e^{-\frac{s}{M^2}}}\, .
\end{eqnarray}

\section{Numerical results and discussions}
The input parameters are taken to be the standard values $\langle
\bar{q}q \rangle=-(0.24\pm 0.01\, \rm{GeV})^3$, $\langle \bar{s}s
\rangle=(0.8\pm 0.2)\langle \bar{q}q \rangle$, $\langle
\bar{s}g_s\sigma G s \rangle=m_0^2\langle \bar{s}s \rangle$,
$m_0^2=(0.8 \pm 0.2)\,\rm{GeV}^2$, $\langle \frac{\alpha_s
GG}{\pi}\rangle=(0.33\,\rm{GeV})^4 $, $m_s=(0.14\pm0.01)\,\rm{GeV}$
 and $m_c=(1.35\pm0.10)\,\rm{GeV}$ at the energy scale  $\mu=1\, \rm{GeV}$
\cite{SVZ79,Reinders85,Ioffe2005}.

 In the conventional QCD sum
rules \cite{SVZ79,Reinders85}, there are two criteria (pole
dominance and convergence of the operator product expansion) for
choosing  the Borel parameter $M^2$ and threshold parameter $s_0$.
\begin{figure}
\centering
\includegraphics[totalheight=7cm,width=8cm]{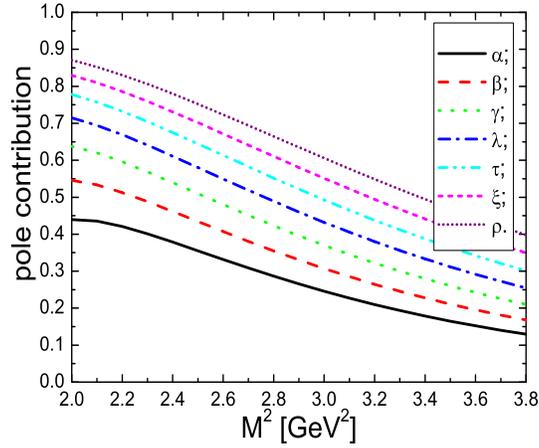}
  \caption{ The contribution from the pole term  with variation of the  Borel parameter $M^2$. The notations
 $\alpha$, $\beta$, $\gamma$, $\lambda$, $\tau$, $\xi$ and $\rho$ correspond to the threshold parameters
 $s_0=19\,\rm{GeV}^2$, $20\,\rm{GeV}^2$,
  $21\,\rm{GeV}^2$, $22\,\rm{GeV}^2$, $23\,\rm{GeV}^2$, $24\,\rm{GeV}^2$ and $25\,\rm{GeV}^2$, respectively.   }
\end{figure}

In Fig.1, we plot the contribution from the pole term with variation
of the threshold parameter $s_0$. From the figure, we can see that
the value $s_0\leq 20 \, \rm{GeV}^2$ is too small to satisfy the
pole dominance condition.

\begin{figure}
 \centering
 \includegraphics[totalheight=5cm,width=6cm]{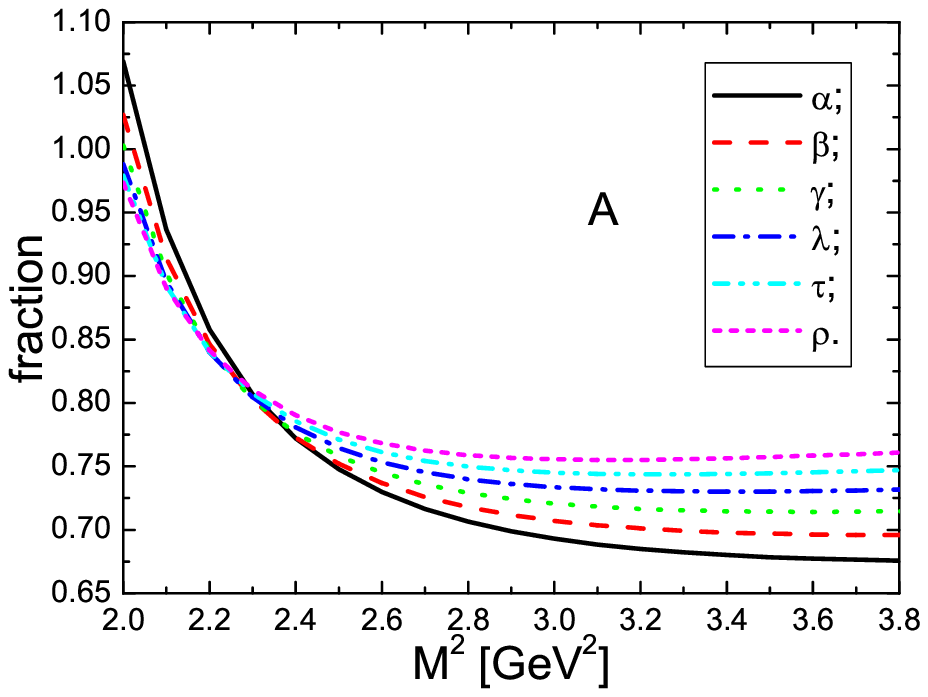}
 \includegraphics[totalheight=5cm,width=6cm]{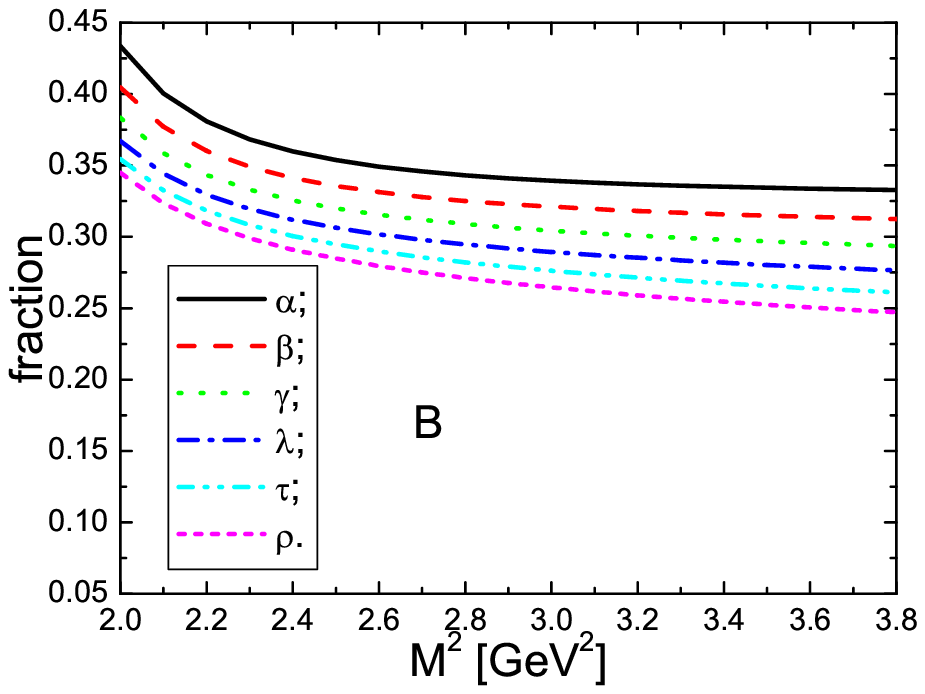}
 \includegraphics[totalheight=5cm,width=6cm]{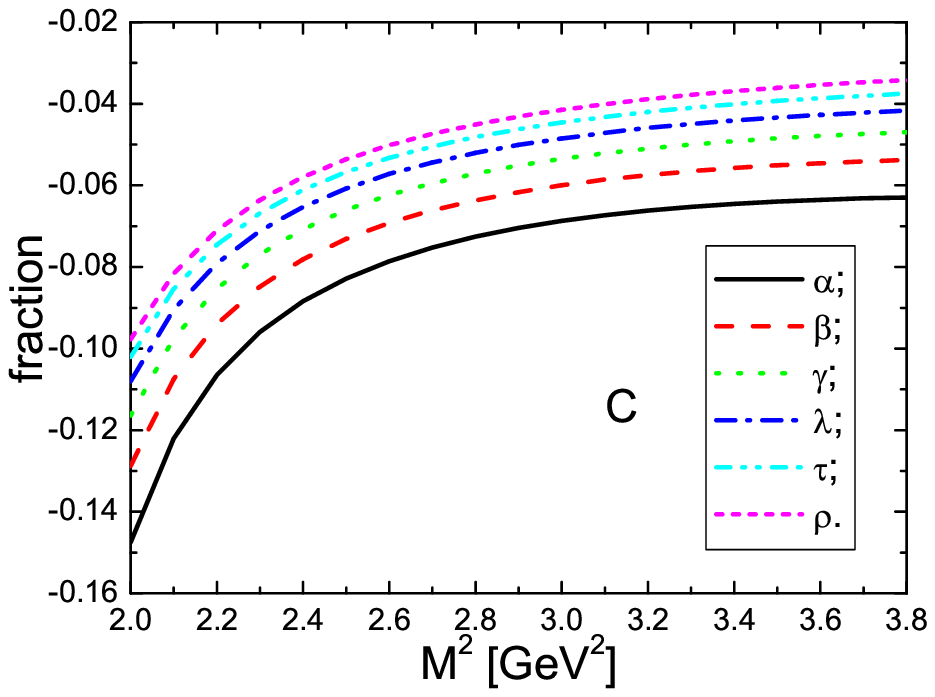}
 \includegraphics[totalheight=5cm,width=6cm]{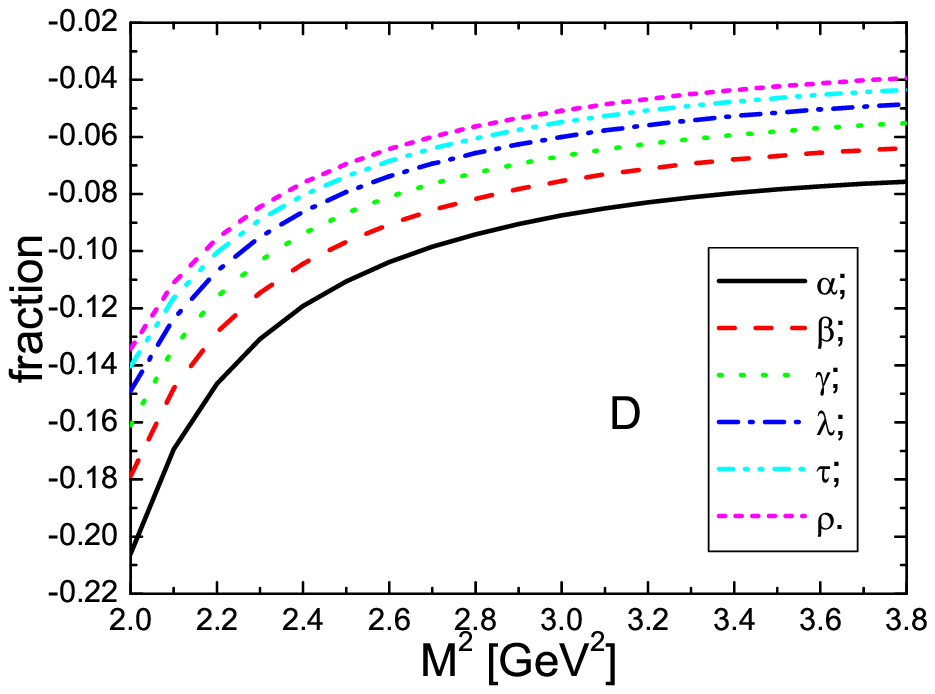}
 \includegraphics[totalheight=5cm,width=6cm]{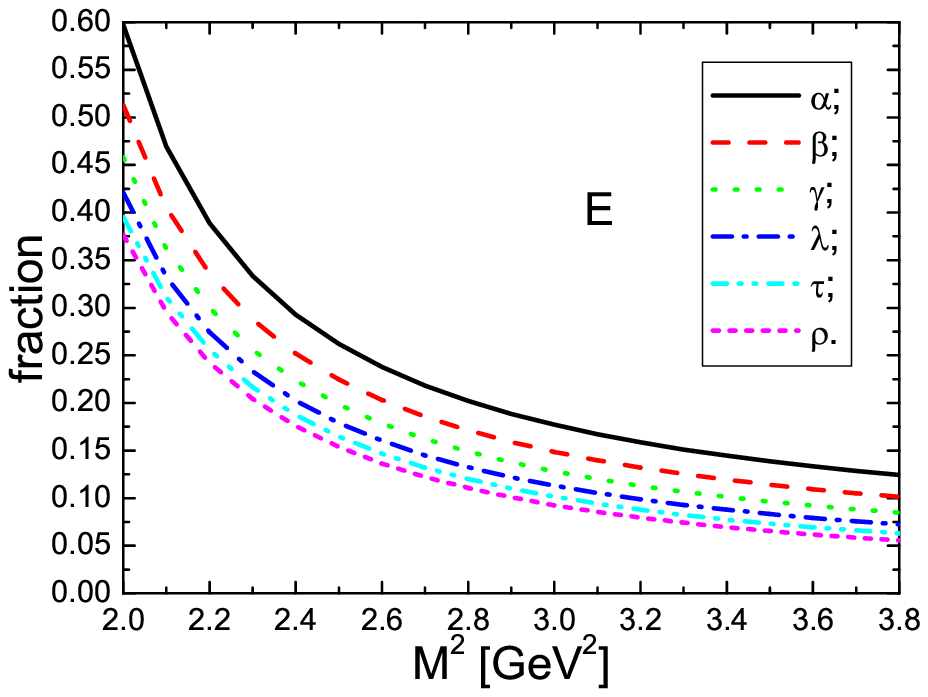}
 \includegraphics[totalheight=5cm,width=6cm]{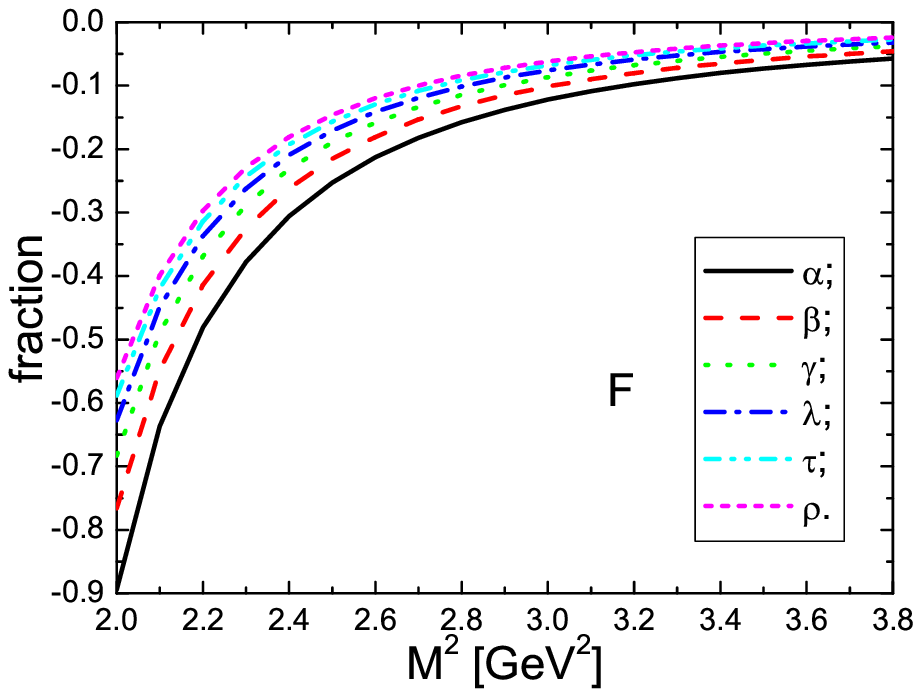}
    \caption{ The contributions from the different terms  with variation of the Borel parameter $M^2$ in the operator product expansion. The $A$,
   $B$, $C$, $D$, $E$ and $F$ correspond to the contributions from
   the perturbative term,
$\langle \bar{s} s \rangle+\langle \bar{s}g_s\sigma G s \rangle$
term,  $\langle \frac{\alpha_s GG}{\pi} \rangle $ term, $\langle
\frac{\alpha_s GG}{\pi} \rangle $+$\langle \frac{\alpha_s GG}{\pi}
\rangle \left[\langle \bar{s} s \rangle +\langle \bar{s}g_s\sigma G
s \rangle+ \langle \bar{s}s \rangle^2\right]$ term, $\langle \bar{s}
s \rangle^2$ term and   $\langle \bar{s} s \rangle\langle
\bar{s}g_s\sigma G s \rangle$+ $\langle \bar{s}g_s\sigma G s
\rangle^2$ term, respectively.     The notations
   $\alpha$, $\beta$, $\gamma$, $\lambda$, $\tau$ and $\rho$ correspond to the threshold
   parameters $s_0=20\,\rm{GeV}^2$,
   $21\,\rm{GeV}^2$, $22\,\rm{GeV}^2$, $23\,\rm{GeV}^2$, $24\,\rm{GeV}^2$ and $25\,\rm{GeV}^2$, respectively.
    Here we take the central values of the input parameters. }
\end{figure}

In Fig.2, we plot the contributions from different terms in the
operator product expansion. The contribution from the term $\langle
\frac{\alpha_s GG}{\pi}\rangle$ is very small, the contributions
from the terms involving the gluon condensates are less than (or
equal) $10\%$  at the values $M^2\geq 2.3\,\rm{GeV}^2$ and $s_0\geq
22\,\rm{GeV}^2$, the gluon condensate plays a minor important role.
The vacuum condensates of the highest dimension $\langle
\bar{s}s\rangle\langle \bar{s} g_s \sigma G s\rangle$+$\langle
\bar{s} g_s \sigma G s\rangle^2$ serve as a criterion for choosing
the Borel parameter $M^2$ and threshold parameter $s_0$. At the
values $M^2_{min} \geq 2.6\,\rm{GeV}^2$ and $s_0\geq
23\,\rm{GeV}^2$, their contributions are less than $15\%$. The
contribution from the vacuum condensate of high dimension $\langle
\bar{s}s\rangle^2$ varies  with the Borel parameter $M^2$ remarkably
and serves as another criterion for choosing the Borel parameter
$M^2$ and threshold parameter $s_0$. At the value $s_0\geq
23\,\rm{GeV}^2$ and $M^2\geq 2.4\,\rm{GeV}^2$, its contribution is
less than (or equal) $ 20 \%$. In the region $M^2\geq
2.6\,\rm{GeV}^2$, the leading contribution comes from the
perturbative term, while the next-to-leading contributions come from
the terms $\langle \bar{s}s\rangle+\langle \bar{s}g_s \sigma
Gs\rangle$. The operator product expansion is convergent at the
values $M^2_{min}\geq 2.6 \, \rm{GeV}^2$ and $s_0 \geq
23\,\rm{GeV}^2$. For the central values of the input parameters, the
contribution from the pole term is larger than  $49\%$ at the values
$M^2_{max} \leq 3.0 \, \rm{GeV}^2 $ and $s_0\geq 23\, \rm{GeV}^2$.

In this article, the threshold   parameter and the Borel parameter
are taken as $s_0=(24 \pm 1)\,\rm{GeV}^2$ and
$M^2=(2.6-3.0)\,\rm{GeV}^2$ respectively, the  contribution from the
pole term is about $(49-72)\%$ for the central values of the other
input parameters, the two criteria of the QCD sum rules are full
filled \cite{SVZ79,Reinders85}. One may expect to take smaller Borel
parameter and threshold parameter to satisfy the two criteria of the
QCD sum rules marginally, however, it is not feasible. The
contributions from the different terms in the operator product
expansion change quickly with variation of the Borel parameter $M^2$
at the value $M^2\leq 2.6\,\rm{GeV}^2$ (see Fig.2) and will not
result in a stable sum rule for the mass (see Fig.3).

Taking into account all uncertainties of the input parameters,
finally we obtain the values of the mass and pole residue of
 the   $Y$, which are  shown in Figs.4-5,
\begin{eqnarray}
M_{Y}&=&(4.43\pm0.16)\,\rm{GeV} \, ,  \nonumber\\
\lambda_{Y}&=&(5.46\pm1.21)\times 10^{-2}\,\rm{GeV}^5 \,   .
\end{eqnarray}
The central value $M_{Y}=4.43\,\rm{GeV}$ is about $200\,\rm{MeV}$
above the $D_s^\ast {\bar D}_s^\ast$ threshold, the $D_s^\ast {\bar
D}_s^\ast$  is probably  a virtual state and not related to the
meson $Y(4140)$. Other possibility, such as a hybrid charmonium is
not excluded. We can explore the hidden charm two-body decay $J/\psi
\phi$ and the open charm two-body decays $D_s \bar{D}_s$, $D_s
\bar{D}^*_s$ to make further studies. More experimental data are
still needed.

From Eq.(11), we can see that the uncertainty of the mass $M_Y$ is
rather small (about $3.6\%$) while the uncertainty of the pole
residue $\lambda_{Y}$ is rather large (about $22.2\%$). The
uncertainties of the input parameters ($\langle \bar{q}q \rangle$,
$\langle \bar{s}s \rangle$, $\langle \bar{s}g_s\sigma G s \rangle$,
 $m_s$ and $m_c$) vary in the range
$(7-25)\%$, so the uncertainty of the pole  residue $\lambda_{Y}$ is
reasonable. We obtain the value of the mass $M_Y$ through a fraction
(see Eq.(10)), the uncertainties in the numerator and denominator
which origin from  a given input parameter (for example, $\langle
\bar{s}s \rangle$, $\langle \bar{s}g_s\sigma G s \rangle$) cancel
out with each other. It is not unexpected that the net uncertainty
is smaller than the uncertainties of the input parameters.

At the energy scale $\mu=1\, \rm{GeV}$, $\frac{\alpha_s}{\pi}\approx
0.19$ \cite{Davier2006}, if the perturbative $\mathcal
{O}(\alpha_s)$ corrections to the perturbative term are companied
with large numerical factors, $1+\xi(s,m_c)\frac{\alpha_s}{\pi}$,
for example, $\xi(s,m_c) >\frac{\pi}{\alpha_s}\approx 5$, the
contributions may be large. We can make a crude estimation by
multiplying the perturbative term  with a numerical factor, say
$1+\xi(s,m_c)\frac{\alpha_s}{\pi}=2$, the mass $M_Y$ decreases
slightly, about $20\, \rm{MeV}$, the  pole residue $\lambda_Y$
increases remarkably. The main contribution comes from the
perturbative term, the large corrections in the numerator and
denominator  cancel out with each other (see Eq.(10)). In fact, the
$\xi(s,m_c)$ are complicated functions of the energy $s$ and the
mass $m_c$, such a crude estimation maybe underestimate the
$\mathcal {O}(\alpha_s)$ corrections, the uncertainties originate
from the $\mathcal {O}(\alpha_s)$ corrections maybe larger.

In this article, we also neglect  the contributions from the
perturbative  corrections $\mathcal {O}(\alpha_s^n)$.  Those
perturbative  corrections can be taken into account in the leading
logarithmic
 approximations through  anomalous dimension factors. After the Borel transform, the effects of those
 corrections are  to multiply each term on the operator product
 expansion side by the factor,
 \begin{eqnarray}
 \left[ \frac{\alpha_s(M^2)}{\alpha_s(\mu^2)}\right]^{2\Gamma_J-\Gamma_{\mathcal
 {O}_n}} \, ,
 \end{eqnarray}
 where the $\Gamma_J$ is the anomalous dimension of the scalar
 interpolating current $J(x)$, the $\Gamma_{\mathcal {O}_n}$ is the anomalous dimension of
 the local operator $\mathcal {O}_n(0)$ in the operator product
 expansion,
 \begin{eqnarray}
 T\left\{J(x)J^{\dagger}(0)\right\}&=&C_n(x) {O}_n(0) \, ,
 \end{eqnarray}
here the $C_n(x)$ is the corresponding Wilson coefficient.

We carry out the operator product expansion at a special energy
scale, say $\mu=1\,\rm{GeV}$, and  can  not smear the scale
dependence by evolving the operator product    expansion side to the
energy scale $M$ through Eq.(12) as the anomalous dimension of the
scalar current $J(x)$ is unknown.
 Furthermore, the anomalous
dimensions of the high dimensional local operators have not been
calculated yet, and their values are poorly known.  In this article,
we set the factor $\left[
\frac{\alpha_s(M^2)}{\alpha_s(\mu^2)}\right]^{2\Gamma_J-\Gamma_{\mathcal
{O}_n}}\approx1$, such an approximation maybe result in some scale
dependence  and  weaken the prediction ability; further studies are
stilled needed.

In the QCD sum rules, the  high dimension vacuum condensates are
always  factorized to lower condensates with vacuum saturation,
  factorization works well in  large $N_c$ limit. In the real world,
  $N_c=3$, there are deviations from the factorable formula, we introduce a factor
  $\kappa$ to parameterize the deviations,
  \begin{eqnarray}
\langle \bar{s} s \rangle^2 \, , \, \langle \bar{s} s \rangle
\langle \bar{s}g_s\sigma G s \rangle \, , \,  \langle
\bar{s}g_s\sigma G s \rangle^2 &\rightarrow& \kappa\langle \bar{s} s
\rangle^2 \, , \, \kappa\langle \bar{s} s \rangle \langle
\bar{s}g_s\sigma G s \rangle \, , \,  \kappa\langle \bar{s}g_s\sigma
G s \rangle^2 \, .
  \end{eqnarray}

In Fig.6, we show the mass $M_Y$ with  variation of the parameter
$\kappa$ at the interval $\kappa=0-2$. From the figure, we can see
that the value of the $M_Y$ changes quickly at the region $M^2\le
2.6\,\rm{GeV}^2$, and  increase with  the $\kappa$ monotonously at
the region $M^2\le 3.2\,\rm{GeV}^2$. At the interval
$M^2=(2.6-3.0)\,\rm{GeV}^2$, the value $\kappa=1\pm1$ leads  to an
uncertainty about $50\,\rm{MeV}$, which is too small to smear the
discrepancy between the present prediction and the experimental
data. In the limit $\kappa=0$, which corresponds to neglecting the
vacuum condensates $\langle \bar{s} s \rangle^2$, $\langle \bar{s} s
\rangle \langle \bar{s}g_s\sigma G s \rangle$ and $\langle
\bar{s}g_s\sigma G s \rangle^2$, we obtain the smallest value.  It
is not unexpected.  From Fig.2E-2F, we can see that there are
cancelations among the  vacuum condensates $\langle \bar{s} s
\rangle^2$, $\langle \bar{s} s \rangle \langle \bar{s}g_s\sigma G s
\rangle$ and $\langle \bar{s}g_s\sigma G s \rangle^2$, the net
contribution is rather small. If we assume the
 $\kappa$ has the typical uncertainty of the QCD sum rules, say about
 $30\%$, the correction   is rather mild, we
 can neglect the uncertainty safely and take $\kappa=1$, i.e. the
 factorization works well.
   In the QCD sum rules for the masses of
the $\rho$ meson and
 nucleon,  $\kappa\geq 1$ \cite{Leinweber97}. If the same value holds
   for the tetraquark states, the deviation from the factorable
   formula means even larger discrepancy between the present
prediction and the experimental data.

The $c$-quark mass appearing in the perturbative terms (see e.g.
Eq.(17)) is usually taken to be the pole mass in the QCD sum rules,
while the choice of the $m_c$ in the leading-order coefficients of
the higher-dimensional terms is arbitrary \cite{Kho9801}.  The
$\overline{MS}$ mass $m_c(m_c^2)$ relates with the pole mass
$\hat{m}$ through the relation
\begin{eqnarray}
m_c(m_c^2) &=&\hat{m}\left[1+\frac{C_F
\alpha_s(m_c^2)}{\pi}+(K-2C_F)\left(\frac{\alpha_s}{\pi}\right)^2+\cdots\right]^{-1}\,
,
\end{eqnarray}
where $K$ depends on the flavor number $n_f$. In this article, we
take the approximation $m_c\approx\hat{m}$ without the $\alpha_s$
corrections for consistency. The value listed in the Particle Data
Group is $m_c(m_c^2)=1.27^{+0.07}_{-0.11} \, \rm{GeV}$ \cite{PDG},
it is reasonable to take the value
$m_c=m_c(1\,\rm{GeV}^2)=(1.35\pm0.10)\,\rm{GeV}$. In Fig.4, we also
present the result with smaller  value $m_c=1.3\,\rm{GeV}$, which
can move down the central value about $0.06\,\rm{GeV}$. The central
value $M_Y=4.37\,\rm{GeV}$ is still larger than the $D_s^\ast {\bar
D}_s^\ast$ threshold about $150\,\rm{MeV}$.

The QCD sum rules  is just a QCD-inspired model, we calculate the
ground state mass   by imposing the two criteria (pole dominance and
convergence of the operator product expansion) of the QCD sum rules.
In fact, we can take smaller threshold parameter $s_0$ and larger
Borel parameter $M^2$ to reproduce the experimental value by
releasing the pole dominance condition.

 We usually consult the experimental data in choosing the
Borel parameter $M^2$ and the threshold parameter $s_0$. The present
experimental knowledge about the phenomenological hadronic spectral
densities of the multiquark states (irrespective of the molecule
type and the diquark-antidiquark type) is  rather vague, even
existence of the multiquark states is not confirmed with confidence.
The nonet scalar mesons below $1\,\rm{GeV}$ (the $f_0(980)$ and
$a_0(980)$ especially) are good candidates for the tetraquark
states. However, they can't satisfy the two criteria of the QCD sum
rules, and result in a reasonable Borel window.  If the perturbative
terms have the main contribution (in the conventional QCD sum rules,
the perturbative terms always have the main contribution), we can
approximate the spectral density with the perturbative term
\cite{Wang0708}, then take the pole dominance condition, and obtain
the approximate  relation,
\begin{eqnarray}
\frac{s_0}{M^2}\geq 4.7 \, .
\end{eqnarray}
If  the Borel parameter has the typical value $M^2=1\, \rm{GeV}^2$,
then $s_0\geq 4.7\, \rm{GeV}^2$, the threshold parameter is too
large for the light tetraquark state candidates $f_0(980)$,
$a_0(980)$, etc.

 Once the
main Fock states of the nonet scalar mesons below $1\,\rm{GeV}^2$
are proved  to be tetraquark states, we can draw the conclusion that
the QCD sum rules are not applicable for the light tetraquark
states. We can either reject the QCD sum rules for the multiquark
states or release one of the two  criteria (pole dominance and
convergence of the operator product expansion) \cite{Wang1}, for
example, we can cut  the threshold parameters for the $f_0(980)$ and
$a_0(980)$ slightly larger than   $1\,\rm{GeV}^2$ by hand.

\begin{figure}
\centering
\includegraphics[totalheight=7cm,width=8cm]{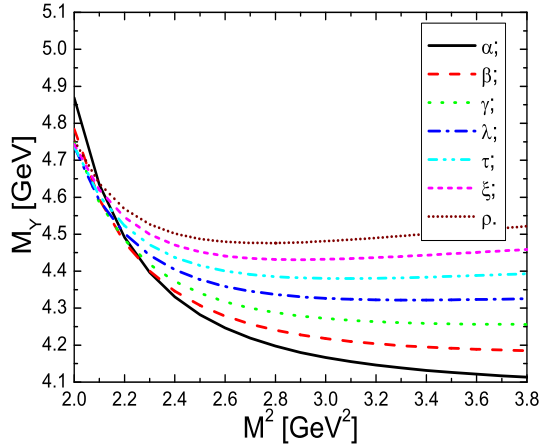}
  \caption{ The mass  with variation of the  Borel parameter $M^2$. The notations
 $\alpha$, $\beta$, $\gamma$, $\lambda$, $\tau$, $\xi$ and $\rho$ correspond to the threshold parameters
 $s_0=19\,\rm{GeV}^2$, $20\,\rm{GeV}^2$,
  $21\,\rm{GeV}^2$, $22\,\rm{GeV}^2$, $23\,\rm{GeV}^2$, $24\,\rm{GeV}^2$ and $25\,\rm{GeV}^2$, respectively.    }
\end{figure}

\begin{figure}
\centering
\includegraphics[totalheight=7cm,width=8cm]{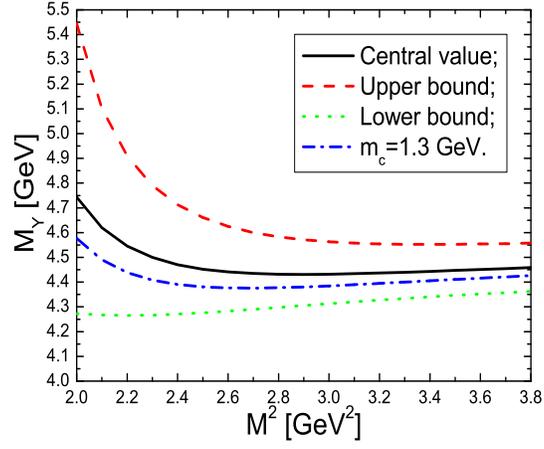}
  \caption{ The mass  with variation of the  Borel parameter $M^2$.    }
\end{figure}

\begin{figure}
\centering
\includegraphics[totalheight=7cm,width=8cm]{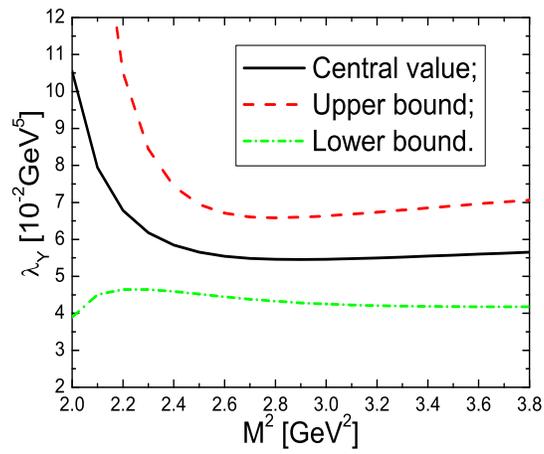}
  \caption{ The  pole residue   with variation of the  Borel parameter $M^2$.    }
\end{figure}

\begin{figure}
\centering
\includegraphics[totalheight=7cm,width=8cm]{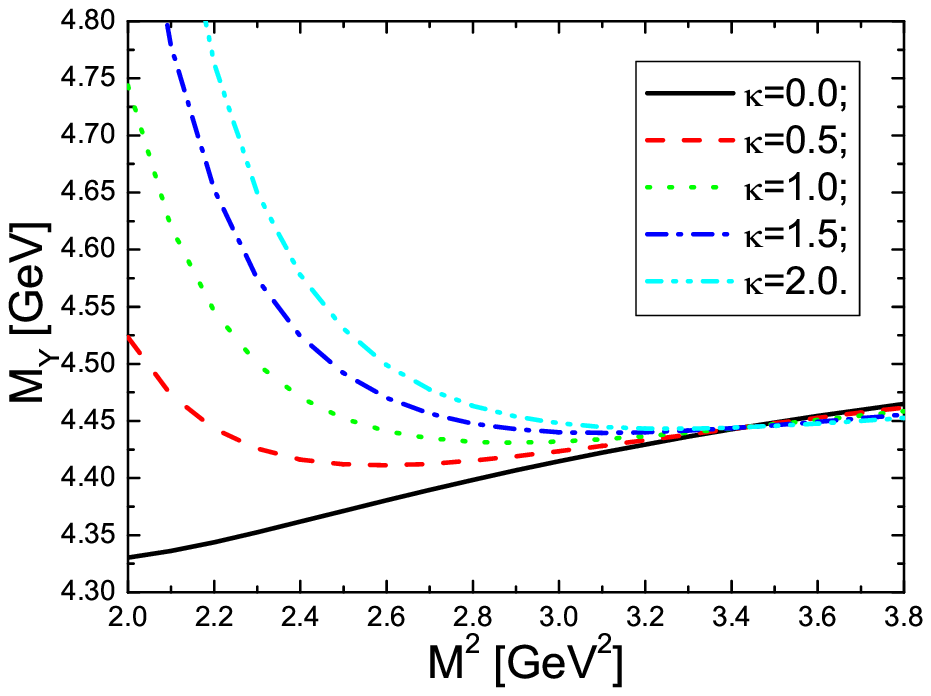}
  \caption{ The mass  with variation of the   parameters $\kappa$ and $M^2$,  other parameters are taken to be  the central values.    }
\end{figure}

\section{Conclusion}
In this article, we assume that there exists a scalar $D_s^\ast
{\bar D}_s^\ast$ molecular state in the $J/\psi \phi$ invariant mass
distribution, and study its mass using the QCD sum rules. Our
predictions depend heavily on  the two criteria (pole dominance and
convergence of the operator product expansion) of the QCD sum rules.
The numerical result indicates that the mass is about
  $M_{Y}=(4.43\pm0.16)\,\rm{GeV}$, which
 is  inconsistent with the experimental data.
    The $D_s^\ast {\bar
D}_s^\ast$  is probably  a virtual state and not related to the
meson $Y(4140)$. Other possibility, such as a hybrid charmonium is
not excluded; more experimental data are still needed to identify
it.

\section*{Appendix}
The spectral densities at the level of the quark-gluon degrees of
freedom:
\begin{eqnarray}
\rho_0(s)&=&\frac{3}{1024 \pi^6}
\int_{\alpha_{i}}^{\alpha_{f}}d\alpha \int_{\beta_{i}}^{1-\alpha}
d\beta
\alpha\beta(1-\alpha-\beta)^3(s-\widetilde{m}^2_c)^2(7s^2-6s\widetilde{m}^2_c+\widetilde{m}^4_c)
\nonumber \\
&&+\frac{3}{1024 \pi^6} \int_{\alpha_{i}}^{\alpha_{f}}d\alpha
\int_{\beta_{i}}^{1-\alpha} d\beta
\alpha\beta(1-\alpha-\beta)^2(s-\widetilde{m}^2_c)^3(3s-\widetilde{m}^2_c)
\nonumber \\
&&+\frac{3m_sm_c}{512 \pi^6} \int_{\alpha_{i}}^{\alpha_{f}}d\alpha
\int_{\beta_{i}}^{1-\alpha} d\beta
(\alpha+\beta)(1-\alpha-\beta)^2(s-\widetilde{m}^2_c)^2(5s-2\widetilde{m}^2_c)
\, ,
\end{eqnarray}

\begin{eqnarray}
\rho_{\langle\bar{s}s\rangle}(s)&=&\frac{3m_s\langle\bar{s}s\rangle}{32
\pi^4} \int_{\alpha_{i}}^{\alpha_{f}}d\alpha
\int_{\beta_{i}}^{1-\alpha} d\beta
\alpha\beta(1-\alpha-\beta)(10s^2-12s\widetilde{m}^2_c+3\widetilde{m}^4_c)
\nonumber \\
&&+\frac{3m_s\langle\bar{s}s\rangle}{32 \pi^4}
\int_{\alpha_{i}}^{\alpha_{f}}d\alpha \int_{\beta_{i}}^{1-\alpha}
d\beta \alpha\beta (s-\widetilde{m}^2_c)(2s-\widetilde{m}^2_c)
\nonumber \\
&&-\frac{m_s\langle\bar{s} g_s \sigma Gs\rangle}{64 \pi^4}
\int_{\alpha_{i}}^{\alpha_{f}}d\alpha \int_{\beta_{i}}^{1-\alpha}
d\beta \alpha\beta
\left[6(2s-\widetilde{m}^2_c)+s^2\delta(s-\widetilde{m}^2_c)\right]
\nonumber \\
&&-\frac{3m_c\langle\bar{s}s\rangle}{32 \pi^4}
\int_{\alpha_{i}}^{\alpha_{f}}d\alpha \int_{\beta_{i}}^{1-\alpha}
d\beta (\alpha+\beta)(1-\alpha-\beta)
(s-\widetilde{m}^2_c)(2s-\widetilde{m}^2_c)
\nonumber \\
&&+\frac{3m_c\langle\bar{s}g_s \sigma Gs\rangle}{128 \pi^4}
\int_{\alpha_{i}}^{\alpha_{f}}d\alpha \int_{\beta_{i}}^{1-\alpha}
d\beta (\alpha+\beta) (3s-2\widetilde{m}^2_c)
\nonumber \\
&&-\frac{3m_sm_c^2\langle\bar{s}s\rangle}{8 \pi^4}
\int_{\alpha_{i}}^{\alpha_{f}}d\alpha \int_{\beta_{i}}^{1-\alpha}
d\beta
(s-\widetilde{m}^2_c)\nonumber \\
&&-\frac{m_s\langle\bar{s}g_s \sigma Gs\rangle}{64 \pi^4}
\int_{\alpha_{i}}^{\alpha_{f}}d\alpha
 \alpha(1-\alpha) (3s-2\widetilde{\widetilde{m}}_c^2)\nonumber \\
&&+\frac{3m_sm_c^2\langle\bar{s}g_s \sigma Gs\rangle}{32 \pi^4}
\int_{\alpha_{i}}^{\alpha_{f}}d\alpha \, ,
\end{eqnarray}

\begin{eqnarray}
\rho_{\langle\bar{s}s\rangle^2}(s)&=&\frac{m_c^2\langle\bar{s}s\rangle^2}{4
\pi^2} \int_{\alpha_{i}}^{\alpha_{f}}d\alpha
+\frac{m_c^2\langle\bar{s}g_s \sigma Gs\rangle^2}{64
\pi^2M^6} \int_{\alpha_{i}}^{\alpha_{f}}d\alpha \widetilde{\widetilde{m}}_c^4 \delta(s-\widetilde{\widetilde{m}}_c^2)\nonumber \\
&&-\frac{m_c^2\langle\bar{s}s\rangle\langle\bar{s}g_s \sigma
Gs\rangle}{8
\pi^2} \int_{\alpha_{i}}^{\alpha_{f}}d\alpha \left[1+\frac{\widetilde{\widetilde{m}}_c^2}{M^2} \right]\delta(s-\widetilde{\widetilde{m}}_c^2)\nonumber \\
&&-\frac{m_sm_c\langle\bar{s}s\rangle^2}{16 \pi^2}
\int_{\alpha_{i}}^{\alpha_{f}}d\alpha
\left[2+s \delta(s-\widetilde{\widetilde{m}}_c^2)\right]\nonumber \\
&&+\frac{5m_sm_c\langle\bar{s}s\rangle\langle\bar{s}g_s \sigma
Gs\rangle}{96\pi^2} \int_{\alpha_{i}}^{\alpha_{f}}d\alpha
\left[1+\frac{\widetilde{\widetilde{m}}_c^2}{M^2}+\frac{\widetilde{\widetilde{m}}_c^4}{2M^4}
\right]\delta(s-\widetilde{\widetilde{m}}_c^2) \, ,
\end{eqnarray}

\begin{eqnarray}
\rho^A_{\langle GG\rangle}(s)&=&-\frac{m_c^2}{256 \pi^4}
\int_{\alpha_{i}}^{\alpha_{f}}d\alpha \int_{\beta_{i}}^{1-\alpha}
d\beta \left(\frac{\alpha}{\beta^2}+\frac{\beta}{\alpha^2}
\right)(1-\alpha-\beta)^3\left[2s-\widetilde{m}^2_c+\frac{s^2}{6}\delta(s-\widetilde{m}^2_c)\right]
\nonumber \\
&&+\frac{3m_sm_c-m_c^2}{512 \pi^4}
\int_{\alpha_{i}}^{\alpha_{f}}d\alpha \int_{\beta_{i}}^{1-\alpha}
d\beta \left(\frac{\alpha}{\beta^2}+\frac{\beta}{\alpha^2}
\right)(1-\alpha-\beta)^2(3s-2\widetilde{m}^2_c)\nonumber \\
&&-\frac{m_sm_c^3}{512 \pi^4} \int_{\alpha_{i}}^{\alpha_{f}}d\alpha
\int_{\beta_{i}}^{1-\alpha} d\beta
\left(\frac{1}{\alpha^3}+\frac{1}{\beta^3}
\right)(\alpha+\beta)(1-\alpha-\beta)^2\left[2
+s\delta(s-\widetilde{m}^2_c)\right]
\nonumber \\
 &&-\frac{1}{512 \pi^4}
\int_{\alpha_{i}}^{\alpha_{f}}d\alpha \int_{\beta_{i}}^{1-\alpha}
d\beta
(\alpha+\beta)(1-\alpha-\beta)^2(10s^2-12s\widetilde{m}^2_c+3\widetilde{m}^4_c)
\nonumber \\
&&+\frac{1}{256 \pi^4} \int_{\alpha_{i}}^{\alpha_{f}}d\alpha
\int_{\beta_{i}}^{1-\alpha} d\beta
(\alpha+\beta)(1-\alpha-\beta)(s-\widetilde{m}^2_c)(2s-\widetilde{m}^2_c)
\nonumber \\
&&-\frac{3m_sm_c}{128 \pi^4} \int_{\alpha_{i}}^{\alpha_{f}}d\alpha
\int_{\beta_{i}}^{1-\alpha} d\beta
(1-\alpha-\beta)(3s-2\widetilde{m}^2_c)
\nonumber \\
&&-\frac{m_sm_c^2\langle\bar{s}s\rangle}{96 \pi^2}
\int_{\alpha_{i}}^{\alpha_{f}}d\alpha \int_{\beta_{i}}^{1-\alpha}
d\beta \left(\frac{\alpha}{\beta^2}+\frac{\beta}{\alpha^2}
\right)(1-\alpha-\beta)\nonumber \\
&&\left[1+\frac{\widetilde{m}^2_c}{M^2}+\frac{\widetilde{m}^4_c}{2M^4}\right]\delta(s-\widetilde{m}^2_c)\nonumber \\
&&-\frac{m_sm_c^2\langle\bar{s}s\rangle}{192\pi^2}
\int_{\alpha_{i}}^{\alpha_{f}}d\alpha \int_{\beta_{i}}^{1-\alpha}
d\beta \left(\frac{\alpha}{\beta^2}+\frac{\beta}{\alpha^2}
\right)\left[1+\frac{\widetilde{m}^2_c}{M^2}\right]\delta(s-\widetilde{m}^2_c)\nonumber \\
&&+\frac{m_sm_c^2\langle\bar{s}g_s \sigma Gs\rangle}{1152\pi^2M^6}
\int_{\alpha_{i}}^{\alpha_{f}}d\alpha \int_{\beta_{i}}^{1-\alpha}
d\beta \left(\frac{\alpha}{\beta^2}+\frac{\beta}{\alpha^2}
\right)\widetilde{m}^4_c\delta(s-\widetilde{m}^2_c)\nonumber \\
&&+\frac{m_sm_c^4\langle\bar{s}s\rangle}{48\pi^2M^2}
\int_{\alpha_{i}}^{\alpha_{f}}d\alpha \int_{\beta_{i}}^{1-\alpha}
d\beta \left(\frac{1}{\alpha^3}+\frac{1}{\beta^3}
\right)\delta(s-\widetilde{m}^2_c)\nonumber \\
&&+\frac{m_c^3\langle\bar{s}s\rangle}{192 \pi^2}
\int_{\alpha_{i}}^{\alpha_{f}}d\alpha \int_{\beta_{i}}^{1-\alpha}
d\beta \left(\frac{1}{\alpha^3}+\frac{1}{\beta^3}
\right)(\alpha+\beta)(1-\alpha-\beta)\nonumber\\
&&\left[1+\frac{\widetilde{m}^2_c}{M^2}
\right]\delta(s-\widetilde{m}^2_c)
\nonumber \\
&&-\frac{m_c^3\langle\bar{s}g_s \sigma Gs\rangle}{768 \pi^2M^4}
\int_{\alpha_{i}}^{\alpha_{f}}d\alpha \int_{\beta_{i}}^{1-\alpha}
d\beta \left(\frac{1}{\alpha^3}+\frac{1}{\beta^3}
\right)(\alpha+\beta)\widetilde{m}^2_c\delta(s-\widetilde{m}^2_c)
\nonumber \\
&&-\frac{m_c\langle\bar{s}s\rangle}{64 \pi^2}
\int_{\alpha_{i}}^{\alpha_{f}}d\alpha \int_{\beta_{i}}^{1-\alpha}
d\beta \left(\frac{\alpha}{\beta^2}+\frac{\beta}{\alpha^2}
\right)(1-\alpha-\beta)\left[2+s\delta(s-\widetilde{m}^2_c)\right]\nonumber \\
&&+\frac{m_c\langle\bar{s}g_s \sigma Gs\rangle}{256 \pi^2}
\int_{\alpha_{i}}^{\alpha_{f}}d\alpha \int_{\beta_{i}}^{1-\alpha}
d\beta \left(\frac{\alpha}{\beta^2}+\frac{\beta}{\alpha^2}
\right)\left[1+\frac{\widetilde{m}^2_c}{M^2}\right]\delta(s-\widetilde{m}^2_c)\nonumber \\
&&-\frac{m_sm_c^2\langle\bar{s}s\rangle}{16 \pi^2}
\int_{\alpha_{i}}^{\alpha_{f}}d\alpha \int_{\beta_{i}}^{1-\alpha}
d\beta \left(\frac{1}{\alpha^2}+\frac{1}{\beta^2}
\right)\delta(s-\widetilde{m}^2_c)\nonumber \\
&&-\frac{m_s\langle\bar{s}s\rangle}{64 \pi^2}
\int_{\alpha_{i}}^{\alpha_{f}}d\alpha \int_{\beta_{i}}^{1-\alpha}
d\beta
(\alpha+\beta)\left[1+\frac{2\widetilde{m}^2_c}{3}\delta(s-\widetilde{m}^2_c)+\frac{\widetilde{m}^4_c}{6M^2}\delta(s-\widetilde{m}^2_c)\right]\nonumber \\
&&+\frac{m_c\langle\bar{s}s\rangle}{32 \pi^2}
\int_{\alpha_{i}}^{\alpha_{f}}d\alpha \int_{\beta_{i}}^{1-\alpha}
d\beta \left[2+s\delta(s-\widetilde{m}^2_c)\right] \, ,
\end{eqnarray}

\begin{eqnarray}
\rho^B_{\langle
GG\rangle}(s)&=&-\frac{m_c^4\langle\bar{s}s\rangle^2}{72M^4}
\int_{\alpha_{i}}^{\alpha_{f}}d\alpha
\left[\frac{1}{\alpha^3}+\frac{1}{(1-\alpha)^3}
\right]\delta(s-\widetilde{\widetilde{m}}_c^2)\nonumber \\
&&-\frac{m_sm_c^4\langle\bar{s}g_s \sigma Gs\rangle}{192\pi^2M^4}
\int_{\alpha_{i}}^{\alpha_{f}}d\alpha
\left[\frac{1}{\alpha^3}+\frac{1}{(1-\alpha)^3}
\right]\delta(s-\widetilde{\widetilde{m}}_c^2)\nonumber \\
&&+\frac{m_sm_c^2\langle\bar{s}g_s \sigma Gs\rangle}{1152\pi^2M^4}
\int_{\alpha_{i}}^{\alpha_{f}}d\alpha
\left[\frac{1-\alpha}{\alpha^2}+\frac{\alpha}{(1-\alpha)^2}
\right]\widetilde{\widetilde{m}}_c^2\delta(s-\widetilde{\widetilde{m}}_c^2)\nonumber \\
&&-\frac{m_sm_c^3\langle\bar{s}s\rangle^2}{288M^4}
\int_{\alpha_{i}}^{\alpha_{f}}d\alpha
\left[\frac{1}{\alpha^3}+\frac{1}{(1-\alpha)^3}
\right]\left[1-\frac{\widetilde{\widetilde{m}}_c^2}{M^2}\right]\delta(s-\widetilde{\widetilde{m}}_c^2)\nonumber \\
&&+\frac{m_c^2\langle\bar{s}s\rangle^2}{24M^2}
\int_{\alpha_{i}}^{\alpha_{f}}d\alpha
\left[\frac{1}{\alpha^2}+\frac{1}{(1-\alpha)^2}
\right]\delta(s-\widetilde{\widetilde{m}}_c^2)\nonumber \\
&&+\frac{m_sm_c^2\langle\bar{s}g_s \sigma Gs\rangle}{64\pi^2M^2}
\int_{\alpha_{i}}^{\alpha_{f}}d\alpha
\left[\frac{1}{\alpha^2}+\frac{1}{(1-\alpha)^2}
\right]\delta(s-\widetilde{\widetilde{m}}_c^2)\nonumber \\
&&-\frac{m_sm_c\langle\bar{s}s\rangle^2}{96M^4}
\int_{\alpha_{i}}^{\alpha_{f}}d\alpha
\left[\frac{1-\alpha}{\alpha^2}+\frac{\alpha}{(1-\alpha)^2}
\right]\widetilde{\widetilde{m}}_c^2\delta(s-\widetilde{\widetilde{m}}_c^2)\nonumber \\
&&+\frac{m_s\langle\bar{s}s\rangle}{384\pi^2}
\int_{\alpha_{i}}^{\alpha_{f}}d\alpha
\left[2+s\delta(s-\widetilde{\widetilde{m}}_c^2)
\right]\nonumber \\
&&-\frac{m_c\langle\bar{s}g_s \sigma Gs\rangle}{128\pi^2}
\int_{\alpha_{i}}^{\alpha_{f}}d\alpha
\left[1+\frac{\widetilde{\widetilde{m}}_c^2}{M^2}\right]\delta(s-\widetilde{\widetilde{m}}_c^2)
\, ,
\end{eqnarray}
where $\alpha_{f}=\frac{1+\sqrt{1-\frac{4m_c^2}{s}}}{2}$,
$\alpha_{i}=\frac{1-\sqrt{1-\frac{4m_c^2}{s}}}{2}$,
$\beta_{i}=\frac{\alpha m_c^2}{\alpha s -m_c^2}$,
$\widetilde{m}_c^2=\frac{(\alpha+\beta)m_c^2}{\alpha\beta}$,
$\widetilde{\widetilde{m}}_c^2=\frac{m_c^2}{\alpha(1-\alpha)}$.

\section*{Acknowledgements}
This  work is supported by National Natural Science Foundation of
China, Grant Number 10775051, and Program for New Century Excellent
Talents in University, Grant Number NCET-07-0282.

\end{document}